\documentclass{du-journals}
\journal{Journal of Holography Applications in Physics}
\ArticleType{Regular article}
\Year{Spring 2027}
\Vol{7}
\No{2}
\Page{\thepage--\pageref{LastPage}}
\Page{xx--xx}
\Received{August 1, 2026}
\Revised{September 17, 2026}
\Accepted{September 19, 2026}
\Doihead{10.22128/jhap.2026.3413.1220}

\title{Joule Thomson expansion of symmergent black holes in extended phase space}
\subtitle{Thermodynamic Structure, Inversion Curves, and Isenthalpic Expansion}

\author[1]{Saheb Soroushfar}
\address[1]{Department of Physics, College of Sciences, Yasuj University,  75918-74934, Yasouj, Iran;\\ E-mail: Soroush@yu.ac.ir}
\author[2]{Sayyed Mehrab Ramezani}
\address[2]{Department of Mathematics, College of Sciences, Yasouj University, Yasouj, Iran.\\ Corresponding Author E-mail: m.ramezani@yu.ac.ir}
\begin{document}

\begin{abstract}
We investigate the Joule Thomson expansion of a symmergent black hole within an extended phase-space formalism in which the thermodynamic pressure is introduced as an independent variable. The symmergent entropy contains the parameter \(\hat{\alpha}\), which modifies the enthalpy and thermodynamic volume relative to their standard forms. We show that the Joule Thomson coefficient depends only on the combination \(8\pi P r_{+}^{2}\), leading to an inversion curve \(P_{\mathrm{i}} = \frac{\pi}{2} T_{\mathrm{i}}^{2}\) that is independent of \(\hat{\alpha}\). The isenthalpic curves, however, retain a parametric dependence on \(\hat{\alpha}\), indicating that the symmergent sector affects the thermodynamic trajectories but not the inversion boundary itself. The inversion pressure and temperature are derived explicitly, and the cooling regime \(P < P_{\mathrm{i}}\) is distinguished from the heating regime \(P > P_{\mathrm{i}}\).
Since the effective geometry is asymptotically AdS-like for positive pressure, the results also admit a holographic reading, in which the Joule Thomson expansion corresponds to an isenthalpic process in the dual field theory.
 The results provide a consistent thermodynamic framework for symmergent black holes and clarify the role of the symmergent parameters in the Joule Thomson process.
\end{abstract}


\newpage
\tableofcontents
\thispagestyle{dups}
\newpage
\section{Introduction}

Black-hole thermodynamics provides a remarkable connection between gravitational dynamics, quantum field theory, and thermodynamic principles. The identification of black-hole entropy with a quantity proportional to the horizon area and the discovery of Hawking radiation established that black holes possess genuine thermodynamic properties \cite{Bek,Haw}. The formulation of the four laws of black-hole mechanics further clarified the close correspondence between horizon dynamics and the laws of thermodynamics \cite{Bard}. Since then, black-hole thermodynamics has been extensively developed within general relativity and a wide range of gravitational theories beyond Einstein gravity \cite{Wald,Sark}.

A significant extension of this framework is provided by the extended phase-space formalism, in which the cosmological constant is promoted to a thermodynamic pressure. In this formulation, the black-hole mass is interpreted as the thermodynamic enthalpy, and the first law acquires an additional pressure volume contribution. This extended thermodynamic interpretation has led to the development of black-hole chemistry and has revealed rich phase structures, including critical behavior analogous to that of ordinary thermodynamic systems \cite{Kast,Dol,Kub,Man}, thereby motivating extensive studies of black-hole thermodynamics in Einstein and modified gravity theories \cite{Sor,Sor1}.
 The resulting framework provides a natural setting for studying thermodynamic processes in which pressure and temperature vary while the enthalpy remains fixed.

One such process is the Joule Thomson expansion. In ordinary thermodynamics, the Joule Thomson process describes an isenthalpic expansion, and the corresponding Joule Thomson coefficient,
\begin{equation}
\mu_{\mathrm{JT}}
=
\left(
\frac{\partial T}{\partial P}
\right)_{H},
\end{equation}
determines the variation of temperature during the expansion. The condition $\mu_{\mathrm{JT}}=0$ defines the inversion curve, which separates the cooling and heating regimes. The application of this concept to black-hole thermodynamics has led to the investigation of inversion curves and isenthalpic trajectories for several black-hole solutions, including charged AdS and Kerr AdS black holes \cite{Ayd,Ayd1}. 
More recently, Joule Thomson processes and the extended thermodynamics of black holes have also been investigated in a variety of modified-gravity settings and matter-coupled gravitational models \cite{CHu,Fat,Kumar}, 
with recent investigations extending the analysis to regular and Schwarzschild-AdS black holes in the extended phase space \cite{Ahmed,Huo}.

Among the theories extending general relativity, symmergent gravity provides a particularly interesting framework in which gravitational dynamics can emerge from quantum effects associated with the gauge sector. The theory was introduced by Demir as a framework in which quadratic-curvature contributions arise through quantum-induced effects in the gravitational sector \cite{Dem1}. The resulting black-hole solutions can possess thermodynamic properties that differ from those of their general-relativistic counterparts. In particular, the thermodynamics and quantum or fluctuation-related corrections of symmergent black holes have recently been investigated in the literature \cite{Bab}.
More generally, the thermodynamic properties of black holes in modified and higher-curvature theories have been studied from several complementary perspectives \cite{Baj,Mish}.
Recent studies have also continued to explore the thermodynamic structure of AdS black holes in generalized gauge-gravity models and modified theories \cite{Sor2}.

Despite these developments, the Joule Thomson expansion of the symmergent black hole considered here has not, to the best of our knowledge, been systematically analyzed within a consistent extended phase-space framework. In particular, the consequences of interpreting the symmergent curvature contribution as a thermodynamic pressure, together with the resulting enthalpy and thermodynamic volume, require a careful formulation before the isenthalpic expansion can be studied. The dependence of the inversion curve and the isenthalpic trajectories on the parameters characterizing the symmergent sector also remains to be clarified.

Black-hole thermodynamics is  naturally connected to holography through
the AdS/CFT correspondence \cite{Mal}, where the thermal properties of an
asymptotically AdS black hole are dual to those of a boundary conformal
field theory. In the extended phase space, varying the thermodynamic pressure
corresponds to varying the central charge of the dual theory \cite{Kub}. For
the symmergent solution considered here, the effective geometry is
asymptotically AdS-like for positive pressure, so the thermodynamic analysis
below can be interpreted within this holographic framework.

In this work, we investigate the Joule Thomson expansion of a static symmergent black hole in the extended phase space. We first establish the geometric framework and express the horizon relation in terms of the symmergent parameters. We then identify the effective thermodynamic pressure associated with the curvature contribution and construct the corresponding thermodynamic quantities. By imposing a consistent extended first law, we derive the thermodynamic enthalpy and volume associated with the black-hole solution. These quantities are subsequently used to obtain the Joule Thomson coefficient and the inversion condition. We analyze the resulting inversion curve and isenthalpic trajectories and examine how the symmergent parameter modifies the cooling and heating regions.

The paper is organized as follows. In Sec.~\ref{sec geometry}, we introduce the symmergent black-hole geometry and derive the horizon relation. The extended phase-space formulation and the corresponding thermodynamic quantities are then established, including the consistent first law, the enthalpy, and the thermodynamic volume. In Sec.~\ref{sec joule}, we derive the Joule Thomson coefficient and obtain the inversion condition. The inversion curve and isenthalpic trajectories are then analyzed, with particular emphasis on the effect of the symmergent parameter. Finally, our conclusions are presented in Sec.~\ref{sec conclusion}.
\section{Geometric and Thermodynamic Framework}
\label{sec geometry}

In this section, we establish the geometric and thermodynamic framework required for the analysis of the Joule Thomson expansion of the symmergent black hole. We begin by recalling the effective gravitational structure of symmergent gravity and introducing the static, spherically symmetric black-hole solution considered in this work. The corresponding horizon relation is then obtained, and the curvature contribution associated with the symmergent sector is incorporated into the extended phase space through an effective thermodynamic pressure. The Hawking temperature and entropy are subsequently determined from the horizon geometry and the gravitational thermodynamic structure. Finally, these quantities are used to formulate a consistent extended first law and to derive the thermodynamic enthalpy and volume. The resulting framework provides the basis for the isenthalpic Joule Thomson expansion studied in Sec.~\ref{sec joule}.

Symmergent gravity is an emergent gravitational framework in which the gravitational sector is connected to quantum effects associated with the underlying matter fields. In its effective description, the induced gravitational dynamics contain the Einstein Hilbert contribution together with curvature corrections generated by the underlying quantum field theory. The coefficient of the quadratic-curvature contribution is controlled by the difference between the numbers of bosonic and fermionic degrees of freedom in the underlying theory. In particular, the corresponding loop coefficient is commonly written as
\begin{equation}
c_{0}
=
\frac{n_{B}-n_{F}}
{128\pi^{2}},
\label{c0definition2}
\end{equation}
where $n_{B}$ and $n_{F}$ denote, respectively, the total numbers of bosonic and fermionic degrees of freedom. The symmergent framework therefore provides a geometric interpretation of the quantum imbalance between these sectors through the curvature scale of the effective gravitational solution \cite{Dem1,Bab}.

The static, spherically symmetric black-hole solution employed in the present work was obtained within this symmergent framework and has been investigated in subsequent studies of the thermodynamic and geometric properties of symmergent black holes \cite{Dem1,Bab}. The corresponding spacetime geometry and its thermodynamic properties are introduced below.
\subsection{Symmergent Black-Hole Geometry}

The static and spherically symmetric black-hole solution employed in the
present work is not derived here but follows directly from the
quadratic-curvature formulation of symmergent gravity developed in
Ref.~\cite{Cim}. For completeness and to establish the notation used in
the thermodynamic analysis, we briefly summarize the main ingredients of
that construction.

In four-dimensional vacuum, the gravitational action of the symmergent
theory can be written in the $f(R)$ form
\begin{equation}
S_{\rm grav}
=
\frac{1}{16\pi G}
\int d^{4}x\sqrt{-g}\,
f(R),
\label{gravaction}
\end{equation}
where
\begin{equation}
f(R)
=
R-\pi G c_{0}R^{2}.
\label{fRsymmergent}
\end{equation}
Equations~\eqref{gravaction} and
\eqref{fRsymmergent} follow directly from the symmergent gravitational
framework presented in Ref.~\cite{Cim}.

The derivative of the curvature function is
\begin{equation}
F(R)
\equiv
\frac{df(R)}{dR}
=
1-2\pi G c_{0}R,
\label{FRsymmergent}
\end{equation}
which is also obtained in Ref.~\cite{Cim}.

Variation of the action with respect to the metric gives the vacuum field
equations
\begin{equation}
F(R)R_{\mu\nu}
-
\frac12 g_{\mu\nu}f(R)
+
\left(
g_{\mu\nu}\Box
-
\nabla_{\mu}\nabla_{\nu}
\right)
F(R)
=
0,
\label{fRfield}
\end{equation}
where
\[
\Box
=
\nabla^{\alpha}\nabla_{\alpha},
\]
and the corresponding traceless form is
\begin{equation}
F(R)R_{\mu\nu}
-
\frac14 g_{\mu\nu}RF(R)
+
\frac14 g_{\mu\nu}\Box F(R)
-
\nabla_{\mu}\nabla_{\nu}F(R)
=
0,
\label{tracelessfield}
\end{equation}
as derived in Ref.~\cite{Cim}.

To obtain the static vacuum solution, one adopts the metric ansatz
\begin{equation}
ds^{2}
=
-B(r)\,dt^{2}
+
\frac{dr^{2}}{B(r)}
+
r^{2}
\left(
d\theta^{2}
+
\sin^{2}\theta\,d\phi^{2}
\right),
\label{metric2}
\end{equation}
whose Ricci scalar is
\begin{equation}
R
=
-
\frac{
r^{2}B''(r)
+
4rB'(r)
+
2B(r)
-
2
}
{r^{2}}.
\label{RicciScalarSymmergent}
\end{equation}
These expressions are identical to those introduced in
Ref.~\cite{Cim}.

Substituting the metric ansatz into the field equations leads to the
independent differential equations
\begin{align}
2FB''+2B'F'-2BF''
-\frac{4}{r}BF'
+\frac{4}{r^{2}}F(1-B)
&=0,
\label{fieldcomponent1}
\\
2FB''+2B'F'+6BF''
-\frac{4}{r}BF'
+\frac{4}{r^{2}}F(1-B)
&=0,
\label{fieldcomponent2}
\\
2FB''+2B'F'+2BF''
-\frac{4}{r}BF'
+\frac{4}{r^{2}}F(1-B)
&=0,
\label{fieldcomponent3}
\end{align}
from which
\begin{equation}
B(r)F''(r)=0
\label{BFsecond}
\end{equation}
is immediately obtained by subtraction.
Outside the horizon, where $B(r)\neq0$, one finds
\begin{equation}
F(R)=a+br.
\label{Flinear}
\end{equation}
Consistency of the field equations further requires
\begin{equation}
b=0,
\end{equation}
so that
\begin{equation}
F(R)=a.
\label{Fconstant}
\end{equation}
These intermediate relations are summarized from the derivation presented
in Ref.~\cite{Cim}.

Integration of the remaining field equation yields
\begin{equation}
B(r)
=
1
+
\frac{C}{r}
+
\frac{a-1}
{24\pi G c_{0}}
r^{2},
\label{Bsolutiongeneral}
\end{equation}
where $C$ is an integration constant.
Identifying
\begin{equation}
C=-2GM,
\end{equation}
and introducing the dimensionless constant
$\hat{\alpha}=a$, the metric function becomes
\begin{equation}
f(r)
=
1
-
\frac{2GM}{r}
-
\frac{1-\hat{\alpha}}
{24\pi G c_{0}}
r^{2}.
\label{metricfunction2}
\end{equation}
Equation~\eqref{metricfunction2} is the exact static black-hole solution
of symmergent gravity obtained in Ref.~\cite{Bab,Cim} and constitutes the
starting point of the present thermodynamic analysis.

Here, $M$ denotes the geometric mass parameter, $G$ is Newton's
gravitational constant, $c_{0}$ characterizes the quadratic-curvature
sector, and $\hat{\alpha}$ is the dimensionless integration constant of
the symmergent solution. The Schwarzschild geometry is recovered in the
limit $\hat{\alpha}\rightarrow1$, where the curvature contribution
vanishes.

The event horizon is determined by the largest positive root of
\begin{equation}
f(r_{+})=0,
\label{horizoncondition2}
\end{equation}
which immediately gives
\begin{equation}
M
=
\frac{r_{+}}{2G}
\left[
1
-
\frac{1-\hat{\alpha}}
{24\pi G c_{0}}
r_{+}^{2}
\right].
\label{massradius2}
\end{equation}

Equation~\eqref{massradius2} expresses the mass parameter in terms of the
event-horizon radius and the symmergent parameters. This relation serves
as the starting point for the extended phase-space formulation developed
in the next subsection, where the curvature contribution is interpreted
as an effective thermodynamic pressure.
\subsection{Extended Phase Space and Thermodynamic Pressure}

The extended phase-space description is obtained by promoting the curvature contribution associated with the symmergent sector to an independent thermodynamic pressure. This construction is analogous to the treatment of the cosmological constant as a thermodynamic pressure in extended black-hole thermodynamics \cite{Kast,Dol,Man}. Accordingly, we define
\begin{equation}
P
=
\frac{\hat{\alpha}-1}
{64\pi^{2}G c_{0}}.
\label{pressdef5}
\end{equation}

With this identification, the curvature term in Eq.~\eqref{metricfunction2} can be rewritten as
\begin{equation}
-
\frac{1-\hat{\alpha}}
{24\pi G c_{0}}\,r^{2}
=
\frac{8\pi P}{3}\,r^{2},
\end{equation}
and the metric function consequently takes the form
\begin{equation}
f(r)
=
1
-
\frac{2GM}{r}
+
\frac{8\pi P}{3}\,r^{2}.
\label{metricpressure2}
\end{equation}
Equation \eqref{metricpressure2} has the standard Schwarzschild–AdS form with an effective
AdS radius $L^2 = \frac{8\pi P}{3}$, so for $ P>0 $ the geometry is asymptotically AdS-like.
The pressure defined in Eq.~\eqref{pressdef5} is treated as an independent thermodynamic variable in the extended phase space. Since $P$ depends on the symmergent curvature parameter $c_{0}$, variations of the thermodynamic pressure correspond to variations of the effective curvature scale associated with the symmergent sector. This identification provides the thermodynamic setting required for studying an isenthalpic Joule Thomson expansion, in which the pressure changes while the enthalpy remains fixed.

Using Eq.~\eqref{metricpressure2}, the horizon condition can be written as
\begin{equation}
1
-
\frac{2GM}{r_{+}}
+
\frac{8\pi P}{3}\,r_{+}^{2}
=
0,
\end{equation}
which yields
\begin{equation}
M
=
\frac{r_{+}}{2G}
\left(
1+
\frac{8\pi P r_{+}^{2}}{3}
\right).
\label{masspressure2}
\end{equation}

The horizon radius therefore provides the natural geometric variable for describing the black-hole state, while the pair $(r_{+},P)$ serves as the basic set of variables for constructing the extended thermodynamic state space. It is important, however, to distinguish the geometric mass parameter $M$ appearing in the metric from the thermodynamic enthalpy $H$. The latter is determined below from the thermodynamic quantities and the consistent extended first law.

\subsection{Thermodynamic Quantities}

The Hawking temperature is determined by the surface gravity associated with the event horizon. For the static metric in Eq.~\eqref{metric2}, the temperature is given by
\begin{equation}
T
=
\frac{f'(r_{+})}{4\pi}.
\label{temperaturedefinition2}
\end{equation}

Using the pressure representation in Eq.~\eqref{metricpressure2} together with the horizon relation, one obtains
\begin{equation}
f'(r_{+})
=
\frac{1}{r_{+}}
+
8\pi P r_{+}.
\label{fprime2}
\end{equation}

Consequently, the Hawking temperature takes the form
\begin{equation}
T
=
\frac{1+8\pi P r_{+}^{2}}
{4\pi r_{+}}.
\label{temp6}
\end{equation}

The entropy of the symmergent black hole is not given by the standard Bekenstein Hawking expression alone. The modified gravitational sector changes the normalization of the gravitational entropy, leading to
\begin{equation}
S
=
\frac{\hat{\alpha}\pi r_{+}^{2}}{G}.
\label{entropy2}
\end{equation}

This expression is the entropy associated with the symmergent black-hole solution and has been obtained in previous studies of its thermodynamic properties \cite{Bab}. The factor $\hat{\alpha}$ represents the modified gravitational normalization and is independent of the symmergent curvature parameter $c_{0}$. Consequently, the entropy differs from the standard geometric area expression by the symmergent normalization factor. Its differential is therefore
\begin{equation}
dS
=
\frac{2\hat{\alpha}\pi r_{+}}{G}\,dr_{+}.
\label{dS2}
\end{equation}

At fixed $G$ and $\hat{\alpha}$, the thermodynamic variables obtained above have the functional dependence
\begin{equation}
T=T(r_{+},P),
\qquad
S=S(r_{+}).
\label{statevariables2}
\end{equation}

These relations specify the temperature and entropy required for the extended thermodynamic description. In particular, the dependence of the temperature on both $r_{+}$ and $P$, together with the purely geometric dependence of the entropy on $r_{+}$, determines the thermodynamic structure needed for the formulation of the extended first law.

\subsection{Consistent Extended First Law}

In the extended phase-space formulation, the thermodynamic potential appropriate to a variable pressure is the enthalpy $H$. The extended first law is therefore written as
\begin{equation}
dH
=
T\,dS
+
V\,dP,
\label{firstlaw2}
\end{equation}
where $V$ denotes the thermodynamic volume. The interpretation of the black-hole mass as an enthalpy in an extended thermodynamic phase space follows the standard formulation of black-hole chemistry \cite{Kast,Dol,Man}.

Using Eqs.~\eqref{temp6} and \eqref{dS2}, the entropy contribution to the first law becomes
\begin{equation}
T\,dS
=
\frac{\hat{\alpha}}{2G}
\left(
1+8\pi P r_{+}^{2}
\right)
dr_{+}.
\label{TdS2}
\end{equation}

Consequently, at fixed pressure,
\begin{equation}
\left(
\frac{\partial H}{\partial r_{+}}
\right)_{P}
=
\frac{\hat{\alpha}}{2G}
\left(
1+8\pi P r_{+}^{2}
\right).
\label{Hr6}
\end{equation}

Integrating this relation with respect to the horizon radius gives
\begin{equation}
H
=
\frac{\hat{\alpha}r_{+}}{2G}
+
\frac{4\hat{\alpha}\pi P r_{+}^{3}}
{3G}
+
h(P),
\label{enthalpyintegration2}
\end{equation}
where $h(P)$ is an integration function depending only on the pressure.

The pressure-dependent integration function represents the freedom in choosing the thermodynamic reference state. We choose the reference state such that this contribution vanishes, namely $h(P)=0$. The enthalpy then becomes
\begin{equation}
H
=
\frac{\hat{\alpha}r_{+}}{2G}
\left(
1+
\frac{8\pi P r_{+}^{2}}{3}
\right).
\label{Hfinal}
\end{equation}

Comparison of Eqs.~\eqref{masspressure2} and \eqref{Hfinal} shows that the thermodynamic enthalpy is related to the geometric mass parameter by
\begin{equation}
H
=
\hat{\alpha}M.
\label{Hmassrelation2}
\end{equation}

Thus, because of the modified symmergent entropy normalization, the thermodynamic enthalpy is not identical to the geometric mass parameter appearing in the metric. This distinction is essential for maintaining consistency between the entropy, the first law, and the thermodynamic volume.

The thermodynamic volume is obtained from the enthalpy according to
\begin{equation}
V
=
\left(
\frac{\partial H}{\partial P}
\right)_{S}.
\label{volume_definition2}
\end{equation}

Since the entropy is a function of $r_{+}$ alone, fixing $S$ is equivalent to fixing $r_{+}$ when $G$ and $\hat{\alpha}$ are held fixed. Therefore,
\begin{equation}
V
=
\left(
\frac{\partial H}{\partial P}
\right)_{r_{+}},
\end{equation}
which yields
\begin{equation}
V
=
\frac{4\hat{\alpha}\pi r_{+}^{3}}
{3G}.
\label{Vfinal}
\end{equation}

The thermodynamic volume is consequently related to the corresponding geometric horizon volume,
\begin{equation}
V_{\mathrm{geo}}
=
\frac{4\pi r_{+}^{3}}{3G},
\end{equation}
through
\begin{equation}
V
=
\hat{\alpha}V_{\mathrm{geo}}.
\label{Vgeometricrelation2}
\end{equation}

The same symmergent normalization factor that modifies the entropy therefore appears in both the enthalpy and the thermodynamic volume. With these definitions, the extended first law is satisfied consistently:
\begin{equation}
dH
=
T\,dS
+
V\,dP.
\label{firstlawfinal2}
\end{equation}

The thermodynamic quantities entering the extended first law are therefore summarized by
\begin{equation}
\begin{aligned}
T
&=
\frac{1+8\pi P r_{+}^{2}}
{4\pi r_{+}},
\\[4pt]
S
&=
\frac{\hat{\alpha}\pi r_{+}^{2}}{G},
\\[4pt]
H
&=
\frac{\hat{\alpha}r_{+}}{2G}
\left(
1+
\frac{8\pi P r_{+}^{2}}{3}
\right),
\\[4pt]
V
&=
\frac{4\hat{\alpha}\pi r_{+}^{3}}
{3G}.
\end{aligned}
\label{thermodynamicsummary2}
\end{equation}

These relations provide the complete geometric and thermodynamic framework required for the subsequent analysis. In particular, the pressure is treated as an independent thermodynamic variable, while the enthalpy in Eq.~\eqref{Hfinal} is held fixed during the Joule Thomson expansion. The resulting Joule Thomson coefficient, inversion condition, and isenthalpic trajectories are investigated in Sec.~\ref{sec joule}.
\section{Joule Thomson Expansion and Inversion Curve}
\label{sec joule}

The Joule Thomson expansion is a throttling process that takes place at
constant enthalpy. In the extended phase-space description of black-hole
thermodynamics, the expansion is therefore analyzed along isenthalpic
trajectories in the pressure temperature plane. The response of the
temperature to a pressure variation along such a trajectory is characterized
by the Joule Thomson coefficient,
\begin{equation}
\mu_{\mathrm{JT}}
=
\left(
\frac{\partial T}{\partial P}
\right)_{H}.
\label{JTcoefficientdef}
\end{equation}

A positive value of $\mu_{\mathrm{JT}}$ corresponds to cooling during a
decrease in pressure, whereas a negative value corresponds to heating. The
boundary separating these two regimes is determined by the inversion
condition $\mu_{\mathrm{JT}}=0$. The resulting set of points in the
pressure temperature plane defines the inversion curve. The Joule Thomson
analysis below therefore consists of determining the coefficient
$\mu_{\mathrm{JT}}$, the associated isenthalpic trajectories, and the
inversion curve.

\subsection{Joule Thomson Coefficient}

The temperature and enthalpy obtained in the previous section are functions
of the horizon radius and pressure, namely $T=T(r_{+},P)$ and
$H=H(r_{+},P)$. During the Joule Thomson expansion, the enthalpy is held
fixed. Since $G$ and $\hat{\alpha}$ are treated as fixed parameters, the
differential of the enthalpy along an isenthalpic trajectory is
\begin{equation}
dH
=
\left(
\frac{\partial H}{\partial r_{+}}
\right)_{P}
dr_{+}
+
\left(
\frac{\partial H}{\partial P}
\right)_{r_{+}}
dP
=
0.
\label{dHcondition}
\end{equation}

Using the enthalpy obtained in the previous section, the required partial
derivatives are
\begin{equation}
\left(
\frac{\partial H}{\partial r_{+}}
\right)_{P}
=
\frac{\hat{\alpha}}{2G}
\left(
1+8\pi P r_{+}^{2}
\right),
\label{Hr6}
\end{equation}
and
\begin{equation}
\left(
\frac{\partial H}{\partial P}
\right)_{r_{+}}
=
\frac{4\hat{\alpha}\pi r_{+}^{3}}
{3G}.
\label{HP6}
\end{equation}

It follows that the variation of the horizon radius with pressure along a
constant-enthalpy trajectory is
\begin{equation}
\left(
\frac{\partial r_{+}}{\partial P}
\right)_{H}
=
-
\frac{
\left(
\frac{\partial H}{\partial P}
\right)_{r_{+}}
}{
\left(
\frac{\partial H}{\partial r_{+}}
\right)_{P}
}
=
-
\frac{
8\pi r_{+}^{3}
}{
3\left(
1+8\pi P r_{+}^{2}
\right)
}.
\label{drhodPfinal}
\end{equation}

The factor $\hat{\alpha}$ cancels in this relation because it appears as a
common multiplicative factor in both enthalpy derivatives. This cancellation
is specific to the constant-enthalpy trajectory and does not imply that
$\hat{\alpha}$ is absent from the thermodynamic quantities themselves.

The Joule Thomson coefficient is defined by Eq.~\eqref{JTcoefficientdef}.
Since the temperature depends on both $P$ and $r_{+}$, its derivative along
an isenthalpic trajectory is obtained from the chain rule:
\begin{equation}
\left(
\frac{\partial T}{\partial P}
\right)_{H}
=
\left(
\frac{\partial T}{\partial P}
\right)_{r_{+}}
+
\left(
\frac{\partial T}{\partial r_{+}}
\right)_{P}
\left(
\frac{\partial r_{+}}{\partial P}
\right)_{H}.
\label{chainrule}
\end{equation}

For
\[
T
=
\frac{1+8\pi P r_{+}^{2}}
{4\pi r_{+}},
\]
the required derivatives are
\begin{equation}
\left(
\frac{\partial T}{\partial P}
\right)_{r_{+}}
=
2r_{+},
\qquad
\left(
\frac{\partial T}{\partial r_{+}}
\right)_{P}
=
2P-\frac{1}{4\pi r_{+}^{2}}.
\label{temperaturederivatives}
\end{equation}

Substituting these expressions together with Eq.~\eqref{drhodPfinal} into
Eq.~\eqref{chainrule}, we obtain
\begin{equation}
\mu_{\mathrm{JT}}
=
2r_{+}
-
\left(
2P-\frac{1}{4\pi r_{+}^{2}}
\right)
\frac{
8\pi r_{+}^{3}
}{
3\left(
1+8\pi P r_{+}^{2}
\right)
}.
\label{jtraw}
\end{equation}

After simplification, the Joule Thomson coefficient takes the form
\begin{equation}
\mu_{\mathrm{JT}}
=
\frac{
r_{+}
\left(
1-8\pi P r_{+}^{2}
\right)
}{
3\left(
1+8\pi P r_{+}^{2}
\right)
}.
\label{jtfinal}
\end{equation}

The explicit dependence on the symmergent normalization factor
$\hat{\alpha}$ has disappeared from $\mu_{\mathrm{JT}}$. This is a direct
consequence of the common multiplicative factor $\hat{\alpha}$ appearing in
the enthalpy derivatives that determine the isenthalpic trajectory. Thus,
although $\hat{\alpha}$ remains present in the entropy, enthalpy, and
thermodynamic volume, it does not affect the Joule Thomson coefficient when
$G$ and $\hat{\alpha}$ are held fixed during the expansion. The resulting
coefficient depends on the state variables through the dimensionless
combination $8\pi P r_{+}^{2}$.


The inversion curve is obtained by imposing
\begin{equation}
\mu_{\mathrm{JT}}=0.
\label{inversioncond6}
\end{equation}

For $r_{+}\neq0$, Eq.~\eqref{jtfinal} immediately gives
\begin{equation}
P_{\mathrm{i}}
=
\frac{1}{8\pi r_{+}^{2}}.
\label{inversionpressure}
\end{equation}

Substituting this relation into the temperature equation yields
\begin{equation}
T_{\mathrm{i}}
=
\frac{1}{2\pi r_{+}}.
\label{inversiontemperature}
\end{equation}

Eliminating $r_{+}$ between the last two equations gives the inversion curve
in the pressure temperature plane:
\begin{equation}
P_{\mathrm{i}}
=
\frac{\pi}{2}
T_{\mathrm{i}}^{2}.
\label{inversioncurve}
\end{equation}
Holographically, the inversion curve \eqref{inversioncurve} separates the cooling regime
\(\mu_{\mathrm{JT}} > 0\) from the heating regime \(\mu_{\mathrm{JT}} < 0\) in
the dual field theory, with the sign of \(\mu_{\mathrm{JT}}\) indicating
whether the dual system cools or heats as the pressure decreases.

Thus, within the thermodynamic framework adopted here, the inversion curve
is independent of $\hat{\alpha}$ when expressed in the $(P,T)$ plane. This
does not mean that the symmergent parameter has no thermodynamic role.
Rather, its explicit contribution cancels from the Joule Thomson coefficient
and, consequently, from the inversion curve, while it remains present in the
underlying entropy, enthalpy, and thermodynamic volume. Its possible effect
on the isenthalpic trajectories is therefore examined separately in the
following subsection.
\subsection{Isenthalpic Curves}

We now analyze the isenthalpic trajectories associated with the
Joule Thomson expansion. In a throttling process, the thermodynamic
enthalpy remains constant. Therefore, each individual trajectory in the
pressure temperature plane is characterized by a fixed value of the
enthalpy. Denoting this constant value by $H_{0}$, Eq.~\eqref{Hfinal}
can be written along a given isenthalpic trajectory as
\begin{equation}
H_{0}
=
\frac{\hat{\alpha}r_{+}}{2G}
\left(
1+
\frac{8\pi P r_{+}^{2}}{3}
\right).
\label{constantH}
\end{equation}

Here, $H_{0}$ is not a new thermodynamic quantity; it represents the
constant value of the enthalpy $H$ along a particular isenthalpic
trajectory. Different values of $H_{0}$ therefore label different
isenthalpic curves.

Solving Eq.~\eqref{constantH} for the pressure gives the parametric
form of the isenthalpic trajectories:
\begin{equation}
P(r_{+};H_{0})
=
\frac{3}{8\pi r_{+}^{2}}
\left(
\frac{2GH_{0}}
{\hat{\alpha}r_{+}}
-1
\right).
\label{Pisenthalpic}
\end{equation}

The corresponding temperature is obtained by substituting this
pressure into the Hawking temperature given in Eq.~\eqref{temp6}:
\begin{equation}
T(r_{+};H_{0})
=
\frac{
1+8\pi P(r_{+};H_{0})r_{+}^{2}
}
{4\pi r_{+}}.
\label{Tisenthalpic}
\end{equation}

Using Eq.~\eqref{Pisenthalpic}, this relation simplifies to
\begin{equation}
T(r_{+};H_{0})
=
\frac{GH_{0}}
{\pi\hat{\alpha}r_{+}^{2}}
-
\frac{1}
{2\pi r_{+}}.
\label{Tisenthalpicfinal}
\end{equation}

Consequently, the isenthalpic trajectories are parametrically
represented by
\begin{equation}
\begin{aligned}
P(r_{+};H_{0})
&=
\frac{3}{8\pi r_{+}^{2}}
\left(
\frac{2GH_{0}}
{\hat{\alpha}r_{+}}
-1
\right),
\\[6pt]
T(r_{+};H_{0})
&=
\frac{GH_{0}}
{\pi\hat{\alpha}r_{+}^{2}}
-
\frac{1}
{2\pi r_{+}}.
\end{aligned}
\label{isenthalpicparam}
\end{equation}

These relations show that, unlike the inversion curve, the isenthalpic
trajectories retain an explicit dependence on the symmergent
normalization factor through the combination
$GH_{0}/\hat{\alpha}$. This distinction is important: the inversion
condition is determined by the vanishing of the Joule Thomson
coefficient and, in the $(P,T)$ representation obtained above, does not
contain $\hat{\alpha}$, whereas the location and shape of a trajectory
with a specified physical enthalpy $H_{0}$ depend on the ratio
$H_{0}/\hat{\alpha}$.

The intersection between a fixed-$H_{0}$ isenthalpic trajectory and the
inversion curve is obtained by equating the pressure on the isenthalpic
trajectory to the inversion pressure at the same horizon radius:
\begin{equation}
P(r_{+};H_{0})
=
P_{\mathrm{i}}(r_{+})
=
\frac{1}{8\pi r_{+}^{2}}.
\label{intersectioncondition}
\end{equation}

Substituting Eq.~\eqref{Pisenthalpic} into this condition gives
\begin{equation}
\frac{3}{8\pi r_{+}^{2}}
\left(
\frac{2GH_{0}}
{\hat{\alpha}r_{+}}
-1
\right)
=
\frac{1}{8\pi r_{+}^{2}},
\end{equation}
or, equivalently,
\begin{equation}
\frac{2GH_{0}}
{\hat{\alpha}r_{+}}
-1
=
\frac{1}{3}.
\end{equation}

The horizon radius at the intersection is therefore
\begin{equation}
r_{+}^{\mathrm{int}}
=
\frac{3GH_{0}}
{2\hat{\alpha}}.
\label{rintersection}
\end{equation}

The corresponding pressure and temperature are obtained by substituting
$r_{+}^{\mathrm{int}}$ into the inversion relations
\eqref{inversionpressure} and \eqref{inversiontemperature}. Thus, each
fixed value of $H_{0}$ determines a specific intersection point with
the inversion curve.

The sign of the Joule Thomson coefficient determines the thermodynamic
behavior along the isenthalpic trajectories. In the physical domain
where
\begin{equation}
1+8\pi P r_{+}^{2}>0,
\label{physicaldomain}
\end{equation}
the result obtained in Eq.~\eqref{jtfinal} gives
\begin{equation}
\mu_{\mathrm{JT}}>0
\quad\Longleftrightarrow\quad
8\pi P r_{+}^{2}<1,
\label{posJT}
\end{equation}
whereas
\begin{equation}
\mu_{\mathrm{JT}}<0
\quad\Longleftrightarrow\quad
8\pi P r_{+}^{2}>1.
\label{negJT}
\end{equation}

Accordingly, the inversion curve separates the cooling and heating
regions of the isenthalpic expansion. The region with
$\mu_{\mathrm{JT}}>0$ corresponds to cooling during a pressure
decrease, whereas the region with $\mu_{\mathrm{JT}}<0$ corresponds to
heating under the same expansion process.

\begin{figure}[h]
    \centering
    \includegraphics[width=0.7\textwidth]{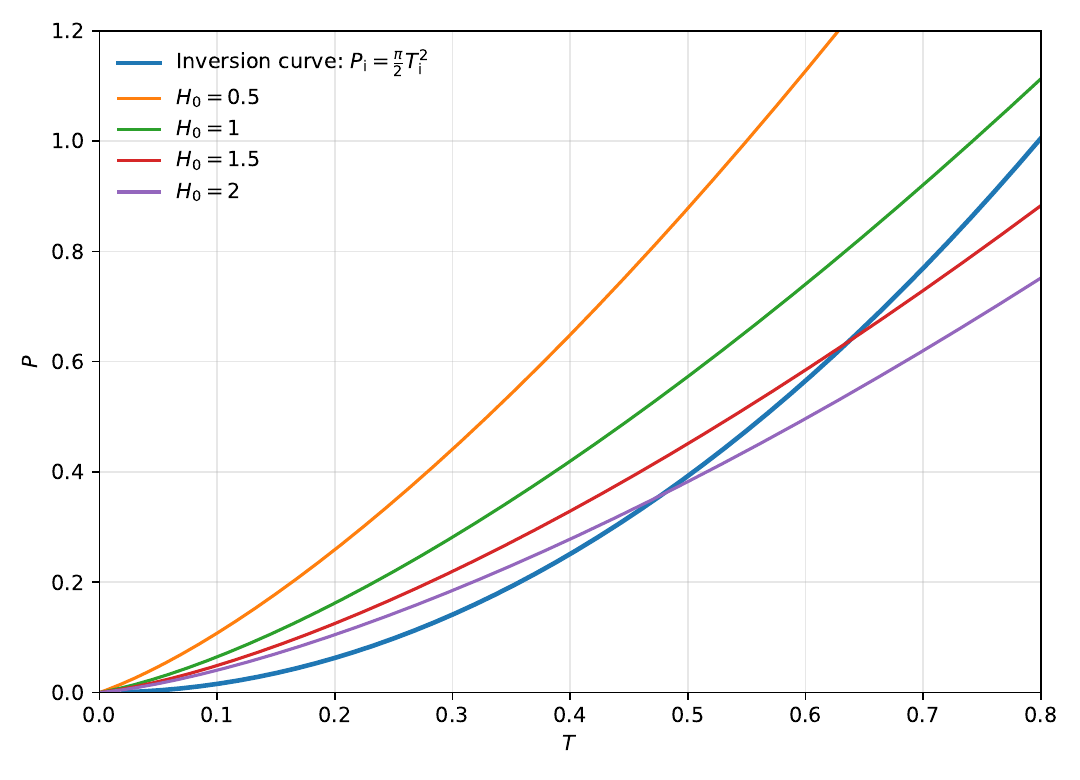}
    \caption{
    The inversion curve and representative isenthalpic trajectories of
    the symmergent black hole in the pressure temperature plane. The
    inversion curve is given by
    $P_{\mathrm{i}}=\frac{\pi}{2}T_{\mathrm{i}}^{2}$, while the remaining
    curves represent trajectories of fixed enthalpy $H_{0}$. The
    intersections of the isenthalpic trajectories with the inversion
    curve correspond to $\mu_{\mathrm{JT}}=0$ and mark the transition
    between the cooling and heating regimes.
    }
    \label{f1}
\end{figure}

Figure~\ref{f1} displays the inversion curve together with several
representative isenthalpic trajectories obtained from
Eq.~\eqref{isenthalpicparam}. The curves are labeled by their fixed
enthalpy values $H_{0}$, while the inversion curve is independent of
this choice. Each intersection between an isenthalpic trajectory and
the inversion curve corresponds to a point at which
$\mu_{\mathrm{JT}}=0$ and therefore marks the transition between the
cooling and heating regimes.

The qualitative structure shown in Fig.~\ref{f1} is consistent with the
standard thermodynamic interpretation of Joule Thomson expansion in
black-hole systems, where isenthalpic trajectories cross the inversion
locus and thereby separate regions of cooling and heating
\cite{Ayd,Ayd1,CHu,Fat}. The present analysis extends this framework to
the symmergent black-hole solution and shows explicitly how the
symmergent normalization factor affects the isenthalpic trajectories
through $H_{0}/\hat{\alpha}$, while it cancels from the inversion
condition and the resulting inversion curve in the $(P,T)$ plane. This
separation between the inversion boundary and the parameter-dependent
isenthalpic paths provides the central thermodynamic structure
underlying the Joule Thomson analysis of the symmergent black hole.
\subsection{Thermodynamic Structure of the Inversion Curve}
\label{subsec:inversionstructure}

The inversion curve obtained in the previous subsection provides the
boundary separating the cooling and heating regimes of the
Joule Thomson expansion. In the present thermodynamic description, its
equation is
\begin{equation}
P_{\mathrm{i}}
=
\frac{\pi}{2}T_{\mathrm{i}}^{2},
\label{inversioncurve3}
\end{equation}
which follows from the condition $\mu_{\mathrm{JT}}=0$ together with the
thermodynamic relations evaluated at the inversion point. Thus,
Eq.~\eqref{inversioncurve3} does not represent an additional
independent thermodynamic condition. Rather, it is the explicit
pressure temperature representation of the inversion condition already
obtained from the Joule Thomson coefficient.

Equation~\eqref{inversioncurve3} shows that the inversion boundary has a
parabolic form in the $(P,T)$ plane. Within the thermodynamic
parametrization adopted here, the resulting curve contains no explicit
dependence on the symmergent normalization factor $\hat{\alpha}$. This
property follows from the cancellation of $\hat{\alpha}$ in the
constant-enthalpy derivative and, consequently, in the expression for
$\mu_{\mathrm{JT}}$. The inversion curve therefore defines a universal
boundary with respect to this normalization factor.

This independence, however, should not be interpreted as indicating
that the complete thermodynamic behavior is independent of
$\hat{\alpha}$. The isenthalpic trajectories retain an explicit
dependence on this parameter through the combination
$GH_{0}/\hat{\alpha}$, as shown by Eq.~\eqref{isenthalpicparam}.
Consequently, the inversion curve and the isenthalpic trajectories
describe two different aspects of the Joule Thomson expansion. The
inversion curve determines the boundary separating the cooling and
heating regimes, whereas the isenthalpic trajectories determine the
actual thermodynamic paths followed by black-hole states with fixed
enthalpy.

The dependence of the inversion pressure and temperature on the horizon
radius follows directly from the inversion relations derived in the
previous subsection:
\begin{equation}
P_{\mathrm{i}}\propto r_{+}^{-2},
\qquad
T_{\mathrm{i}}\propto r_{+}^{-1}.
\label{inversionscaling}
\end{equation}

Accordingly, a decrease in the horizon radius leads to an increase in
both the inversion pressure and the inversion temperature. Conversely,
larger horizon radii correspond to lower values of these two
quantities. This behavior originates from the inversion condition, which
fixes the dimensionless combination $8\pi P r_{+}^{2}$ at the inversion
point. The scaling relations also demonstrate that the inversion
properties are directly connected to the horizon geometry, even though
the final pressure temperature representation in
Eq.~\eqref{inversioncurve3} does not contain an explicit factor of
$\hat{\alpha}$.

The cooling and heating regimes can be identified by comparing the
thermodynamic pressure with the inversion pressure at a given horizon
radius. In the physical domain
\begin{equation}
1+8\pi P r_{+}^{2}>0,
\label{physicaldomain3}
\end{equation}
the sign of the Joule Thomson coefficient is determined by the relative
magnitude of $P$ and $P_{\mathrm{i}}$. Therefore,
\begin{equation}
P<P_{\mathrm{i}}
\quad\Longrightarrow\quad
\mu_{\mathrm{JT}}>0,
\label{coolingcondition3}
\end{equation}
corresponds to the cooling regime, whereas
\begin{equation}
P>P_{\mathrm{i}}
\quad\Longrightarrow\quad
\mu_{\mathrm{JT}}<0,
\label{heatingcondition3}
\end{equation}
corresponds to the heating regime. At $P=P_{\mathrm{i}}$, the
Joule Thomson coefficient vanishes and the system lies on the
inversion boundary. Thus, during a pressure decrease along an
isenthalpic trajectory, the region with $\mu_{\mathrm{JT}}>0$
corresponds to cooling, whereas the region with
$\mu_{\mathrm{JT}}<0$ corresponds to heating.

The thermodynamic regimes can therefore be summarized as
\begin{equation}
\begin{cases}
P<P_{\mathrm{i}},
&
\mu_{\mathrm{JT}}>0
\quad\text{(cooling)},\\[4pt]
P=P_{\mathrm{i}},
&
\mu_{\mathrm{JT}}=0
\quad\text{(inversion point)},\\[4pt]
P>P_{\mathrm{i}},
&
\mu_{\mathrm{JT}}<0
\quad\text{(heating)}.
\end{cases}
\label{regimesummary3}
\end{equation}

\begin{figure}[h]
  \centering
  \includegraphics[width=0.7\textwidth]{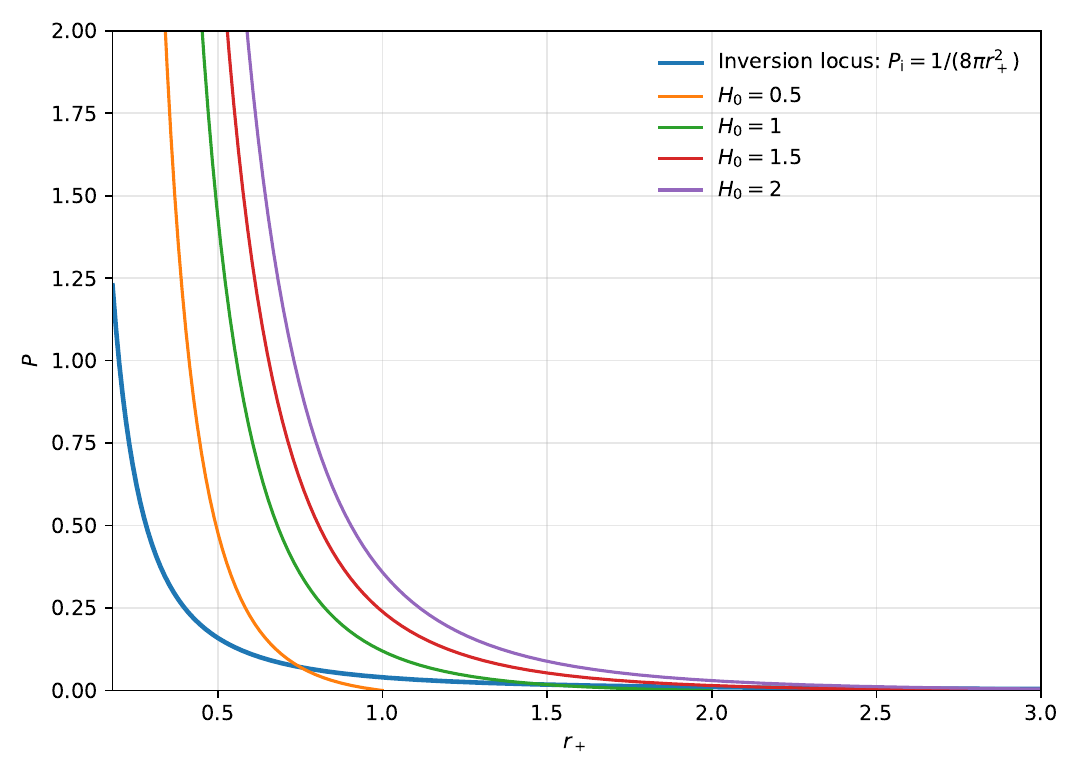}
  \caption{
  The inversion locus and representative isenthalpic trajectories in
  the $(P,r_{+})$ plane. The inversion locus is given by
  $P_{\mathrm{i}}=1/(8\pi r_{+}^{2})$, while the remaining curves
  correspond to different fixed enthalpy values. Their intersections
  identify the horizon radii at which $\mu_{\mathrm{JT}}=0$ and mark the
  transition between the cooling and heating regimes.
  }
  \label{f2}
\end{figure}

The relation between the inversion locus and the constant-enthalpy
trajectories is shown in Fig.~\ref{f2}. The figure is plotted in the
$(P,r_{+})$ plane. The inversion locus is determined by
\begin{equation}
P_{\mathrm{i}}
=
\frac{1}{8\pi r_{+}^{2}},
\end{equation}
whereas the remaining curves represent isenthalpic trajectories
corresponding to different fixed values of the enthalpy. Their
intersections with the inversion locus therefore identify the horizon
radii at which the corresponding thermodynamic states satisfy
$\mu_{\mathrm{JT}}=0$.

Figure~\ref{f2} provides a direct geometric representation of the
different roles played by the inversion locus and the isenthalpic
trajectories. The inversion locus decreases monotonically as the horizon
radius increases, in accordance with the scaling
$P_{\mathrm{i}}\propto r_{+}^{-2}$. It is therefore located at higher
pressures for smaller black holes and approaches the low-pressure region
as $r_{+}$ becomes larger. This behavior is distinct from that of the
isenthalpic trajectories, whose positions depend on the chosen value of
the enthalpy.

For the representative enthalpy values displayed in the figure, the
curves are systematically displaced as the fixed enthalpy is changed.
At a given pressure, trajectories with larger values of the enthalpy
extend toward larger horizon radii. This behavior follows directly from
the fixed-enthalpy parametrization and shows that the location of a
thermodynamic path in the $(P,r_{+})$ plane is not determined solely by
the inversion condition.

Each representative isenthalpic trajectory intersects the inversion
locus at a distinct point. These intersections correspond precisely to
the states for which
$\mu_{\mathrm{JT}}=0$. They therefore mark the points at which a
constant-enthalpy trajectory reaches the boundary separating the two
Joule Thomson regimes. The intersection radius depends on the fixed
enthalpy and, through the combination $H_{0}/\hat{\alpha}$, on the
symmergent normalization factor.

The figure consequently demonstrates an important distinction between
the inversion locus and the isenthalpic curves. The inversion locus is
fixed by the condition for vanishing Joule Thomson coefficient and has
a universal form within the present thermodynamic description. The
isenthalpic trajectories, in contrast, depend on the thermodynamic
enthalpy and on the symmergent normalization through their
parametrization. Hence, different black holes with different fixed
enthalpies follow different paths in the $(P,r_{+})$ plane, although
they encounter the same inversion locus.

At the intersection points, the Joule Thomson coefficient changes sign.
The side of the inversion locus corresponding to
$\mu_{\mathrm{JT}}>0$ represents the cooling regime during a pressure
decrease, whereas the region with $\mu_{\mathrm{JT}}<0$ represents the
heating regime. The figure therefore does not merely display the
inversion relation and a set of constant-enthalpy curves; it also
illustrates how individual isenthalpic thermodynamic paths encounter
the universal boundary and undergo the transition between the two
Joule Thomson behaviors.
\section{ Conclusion}
\label{sec conclusion}

In this work, we have studied the Joule Thomson expansion of a symmergent black hole within a consistent extended phase-space formalism. By identifying the symmergent curvature contribution with a thermodynamic pressure, we constructed the enthalpy, temperature, entropy, and thermodynamic volume from the horizon geometry and the symmergent parameters.

We derived the Joule Thomson coefficient and found that it is independent of the symmergent normalization factor \(\hat{\alpha}\). This cancellation arises from the consistent first-law formulation, in which the same factor appears in both the enthalpy and the volume derivatives. Consequently, the inversion curve \(P_{\mathrm{i}} = \frac{\pi}{2} T_{\mathrm{i}}^{2}\) is also independent of \(\hat{\alpha}\).

In contrast, the isenthalpic trajectories retain a parametric dependence on \(\hat{\alpha}\). This shows that while the inversion boundary is universal with respect to the symmergent normalization, the thermodynamic paths followed during the expansion are modified by the symmergent sector. We also derived the intersection of the isenthalpic curves with the inversion locus and identified the horizon radius at which the Joule Thomson coefficient changes sign.

The cooling regime \(P < P_{\mathrm{i}}\) and the heating regime \(P > P_{\mathrm{i}}\) were distinguished explicitly. The results provide a complete thermodynamic description of the Joule Thomson expansion for the symmergent black hole and clarify the role of the symmergent parameters in the isenthalpic process.

The AdS-like form of the effective geometry for positive pressure places this analysis in a holographic setting. In the dual picture, the
Joule--Thomson expansion is an isenthalpic process and the inversion curve separates cooling from heating. A more detailed holographic study of the symmergent parameter \(\hat{\alpha}\) would be a natural next step.

Several directions remain open for future work. It would be interesting to extend this analysis to rotating or charged symmergent black holes, where additional conserved charges could enrich the thermodynamic structure. The inclusion of quantum or fluctuation corrections to the entropy could also provide further insights. Finally, the connection between the Joule Thomson expansion and other thermodynamic processes in modified gravity frameworks deserves further investigation.

\end{document}